\documentclass[preprint,aps]{revtex4}

\usepackage{graphicx}
\usepackage{dcolumn}

\begin{document}

\preprint{Lebed-Wu-PRL}

\title{Soliton Wall Superlattice in
Quasi-One-Dimensional Conductor
(Per)$_2$Pt(mnt)$_2$}

\author{A.G. Lebed$^*$}
\author{Si Wu}

\affiliation{Department of Physics, University of Arizona, 1118 E.
4-th Street, Tucson, AZ 85721, USA}

\begin{abstract}
We suggest a model to explain the appearance of a high resistance
high magnetic field charge-density-wave (CDW) phase, discovered by
D. Graf et al. [Phys. Rev. Lett. \underline{93}, 076406 (2004)] in
(Per)$_2$Pt(mnt)$_2$. In particular, we show that the Pauli
spin-splitting effects improve the nesting properties of a realistic
quasi-one-dimensional electron spectrum and, therefore, a high
resistance Peierls CDW phase is stabilized in high magnetic fields.
In low and very high magnetic fields, a periodic soliton
wall superlattice (SWS) phase is found to be a ground state. We
suggest experimental studies of the predicted phase transitions
between the Peierls and SWS CDW phases in (Per)$_2$Pt(mnt)$_2$ to
discover a unique SWS phase.
 \\ \\ PACS numbers: 71.45.Lr, 74.70.Kn, 71.10.-w
\end{abstract}

\maketitle

\pagebreak

It is well known that the charge-density-wave (CDW) phases are
destroyed by the Pauli spin-splitting effects in a magnetic field
[1-6], whereas  the spin-density-wave (SDW) phases are not sensitive
to them [2,7-12]. 
Moreover, it is demonstrated both theoretically [2,7-9,12] and 
experimentally [10-12] that the SDW phases can be generated 
by the orbitals effects in a magnetic field due to the so-called field 
induced dimensional crossovers (FIDC) [2,12]. 
An idea about the FIDC has been applied to the CDW phases [3,4], 
where they can also be restored by the orbital effects in a magnetic 
field at very low temperatures [4]. 
Therefore, the recent remarkable discovery of the high resistive high 
magnetic field CDW phases in (Per)$_2$X(mnt)$_2$ ($X= Pt$ and $Au$)
quasi-one-dimensional (Q1D) conductors by Graf et al. [13] has been
tentatively prescribed [13-15] to the FIDC effects [2-4,12]. 
This explanation may be adequate to some degree only in the case of 
$X=Au$, where the high resistance CDW phase occurs only for a magnetic
field, perpendicular to the conducting chains, ${\bf H} \parallel
{\bf c}$ and ${\bf H} \parallel {\bf a}$ (see discussion in the 
end of the Letter).

On the other hand, the high resistance high magnetic field CDW phase
in (Per)$_2$Pt(mnt)$_2$ conductor is shown [13,14] to appear at any
direction of a magnetic field. 
In particular, this phase exists for a magnetic field, parallel to the conducting 
chains, ${\bf H} \parallel {\bf b}$, where the orbital effects are absent and, thus,
the FIDC effects  [2-4,12] do 
not occur. 
Independence of the main features of the CDW phase diagram in 
(Per)$_2$Pt(mnt)$_2$ conductor on a magnetic field direction indicates 
that the high resistance CDW phase in this compound 
is unique and cannot be described by the previous theories 
[1-4,7-9,12].

The main goals of our Letter are as follows. Firstly, we suggest a
theoretical approach, based on a realistic model for a Q1D
electron spectrum, to describe the main properties of the CDW
phase diagram in (Per)$_2$Pt(mnt)$_2$. 
In particular, we demonstrate that the Pauli  spin-splitting effects in 
a magnetic field improve the nesting properties of the Q1D 
electron spectrum. 
This stabilizes a textbook high resistance Peierls 
phase in high arbitrary directed magnetic fields, in contrast to the 
previous theories.
[Below, we call this phenomenon spin improved nesting (SINe)].

Secondly, we show that, in low and very high magnetic
fields, a unique CDW phase - a soliton wall superlattice (SWS)
state - has to appear. 
This semiconducting phase with two energy gaps is characterized 
by a periodically arranged soliton and anti-soliton walls with distance 
between them and values of the energy gaps being functions 
of a magnetic field. 
We predict that phase transitions occur between the 
conventional Peierls and unconventional SWS phases and suggest 
to study them to discover a unique SWS phase in 
(Per)$_2$Pt(mnt)$_2$.
It is important that the existing experimental data on an activated 
behavior of resistivity [13]  are in agreement with the Peierls-SWS 
phase transitions scenario, suggested in the Letter.

Let us discuss the SINe phenomenon, which results in a stabilization 
of the Peierls CDW phase in high magnetic fields, using qualitative 
arguments.
Below, we accept a simplified Q1D electron spectrum of
(Per)$_2$Pt(mnt)$_2$ conductor with four plane sheets of the Fermi
surface (FS) [16],
\begin{equation}
\varepsilon_{\alpha}^{\pm}({\bf p})=\pm v_F\big[p_x\mp p_F \pm
(\Delta p/2)(-1)^{\alpha}\big],
\end{equation}
where +(-) stands for right (left) part of the FS; $p_F$ and $v_F$ are
the average Fermi momentum and the Fermi velocity, $\alpha=1(2)$
stands for the first (second) perylene conducting chain [16], $\Delta p$
is a difference between the values of the Fermi momenta on two 
conducting chains. 
(We note that such kind of an electron spectrum with four slightly 
corrugated sheeets of the FS has been suggested on a basis of the 
band calculations [17], experimentally observed quantum interference 
oscillations [18], and Landau levels 
quantization [19]).

In a magnetic field, the electron spectrum (1) is split into eight
plane sheets of the FS,
\begin{equation}
\varepsilon_{\alpha\sigma}^{\pm}({\bf p})=\pm v_F\big[p_x\mp p_F \pm
(\Delta p/2)(-1)^{\alpha}\big]-\sigma\mu_BH \ ,
\end{equation}
where $\sigma=+1(-1)$ stands for spin up (down), $\mu_B$ is the Bohr
magneton (see Fig.1). As seen from Fig.1, there exist four
competing nesting vectors, $Q_{1,+1}$, $Q_{1,-1}$, $Q_{2,+1}$, and $Q_{2,-1}$, 
for the CDW instability, which pair electrons near $+p_F$ with spin up (down)
and holes near $-p_F$ with spin up (down).

It is natural that four nesting vectors,
\begin{equation}
Q_{\alpha\sigma}=2p_F+q_{\alpha\sigma},\ \ \ \ 
q_{\alpha\sigma}= (-1)^{\alpha} \Delta p  -2\sigma\mu_BH/v_F,
\end{equation}
may correspond to several energy gaps in an electron spectrum 
at high values of the parameters $\Delta p$ and 
$2\mu_BH/v_F$. 
As we show below, this results in the appearance of the SWS phase
with two energy gaps [20] (see Fig.2), which is in a qualitative agreement 
with a general theory of solitons and soliton superstructures 
[21-24].

Nevertheless, as seen from Eq.(3) and Fig.1, at some critical 
magnetic field,
\begin{equation}
H^*=\Delta pv_F/2\mu_B \ ,
\end{equation}
two nesting vectors coincide, 
$Q_{1,-1}=Q_{2,+1}=2p_F$.
Therefore, in the vicinity of this critical field, $H \approx H_p^*$, 
the Peierls CDW phase with $Q=2p_F$ and one gap in an electron 
spectrum becomes more stable than the SWS one 
(see Figs.2,3). 
In other words, at $H\approx H_p^*$, the Pauli spin-splitting effects 
improve nesting properties of the electron spectrum (2), which stabilizes 
a textbook Peierls phase with the nesting vector,
\begin{equation}
{\bf Q}=(2p_F,0,0) \ .
\end{equation}
We suggest that this SINe phenomenon is responsible for the
experimental stabilization of the high resistance high magnetic field
phase in (Per)$_2$Pt(mnt)$_2$ 
[13-15].

Let us consider a formation of the CDW phase, corresponding to
the nesting vector,
\begin{equation}
{\bf Q}=(2p_F+q,0,0),
\end{equation}
where the CDW order parameter is
\begin{equation}
\Delta_{CDW}(x)= \Delta_q{\rm e}^{i(2p_F+q)x} + \Delta^*_q{\rm e}^{-i(2p_F+q)x} \ ,
\end{equation}
using Green functions methods [25].
In this case, a mean field Hamiltonian of the electrons interacting
with the crystalline lattice can be written as,
\begin{eqnarray}
\hat H &=&\sum_{\alpha=1,2}\sum_{\sigma= \pm 1}\sum_{\xi}\Big\{
a_{\alpha\sigma}^{\dagger}(\xi)a_{\alpha\sigma}(\xi)
\big[\varepsilon_{\alpha\sigma}^+(\xi)-\mu\big]+
b_{\alpha\sigma}^{\dagger}(\xi)b_{\alpha\sigma}(\xi)
\big[\varepsilon_{\alpha\sigma}^-(\xi)-\mu\big]\Big\}\nonumber\\
&+&\sum_{\alpha=1,2}\sum_{\sigma= \pm 1}\sum_{\xi}\Big\{
a_{\alpha\sigma}^{\dagger}(\xi)b_{\alpha\sigma}(\xi-q) \Delta_q+
b_{\alpha\sigma}^{\dagger}(\xi)a_{\alpha\sigma}(\xi+q)\Delta_q^*\Big\},
\end{eqnarray}
where 
\begin{equation}
\Psi_{\alpha \sigma} (x)= \exp(-i p_Fx) \sum_{\xi} e^{i \xi x} b_{\alpha \sigma}(\xi) 
+ \exp(i p_Fx) \sum_{\xi} e^{i \xi x} a_{\alpha \sigma} (\xi) 
\end{equation}
is a field operator, $a_{\alpha \sigma} (\xi)$ and $b_{\alpha \sigma}(\xi)$
are electron annihilation operators near right and left sheets of the Q1D FS
(1), correspondingly.

Using standard definitions of the normal and anomalous (Gor'kov)
Green functions,
\begin{equation}
G_{\alpha\sigma}^{++}(\xi, \tau)=-\langle
T_{\tau}a_{\alpha\sigma}(\xi,\tau)
a_{\alpha\sigma}^{\dagger}(\xi,0)\rangle,\ \ \
G_{\alpha\sigma}^{-+}(\xi, \tau)=-\langle
T_{\tau}b_{\alpha\sigma}(\xi-q,\tau)
a_{\alpha\sigma}^{\dagger}(\xi,0)\rangle,
\end{equation}
we find that the Green functions obey the following equations,
\begin{equation}
\Big(i\omega_n-\big[\varepsilon_{\alpha\sigma}^+(\xi)-\mu\big]\Big)
G_{\alpha\sigma}^{++}(\xi,i\omega_n)-\Delta_q
G_{\alpha\sigma}^{-+}(\xi,i\omega_n)=1,
\end{equation}
\begin{equation}
\Big(i\omega_n-\big[\varepsilon_{\alpha\sigma}^-(\xi-q)-\mu\big]\Big)
G_{\alpha\sigma}^{-+}(\xi,i\omega_n)-\Delta^*_q
G_{\alpha\sigma}^{++}(\xi,i\omega_n)=0,
\end{equation}
where the self-consistent gap equation is
\begin{equation}
\Delta^*_q=-g^2\sum_{\alpha=1,2}\sum_{\sigma= \pm 1}\sum_{\xi}T\sum_{\omega_n}
G_{\alpha\sigma}^{-+}(\xi,i\omega_n),
\end{equation}
with $\omega_n = 2 \pi T (n + \frac{1}{2})$ being the Matsubara 
frequency.

Below, we are interested in a phase transition line between the metallic
and CDW phases, therefore, we need to solve the linearized 
Eqs.(11)-(13).
As a result, we find the following expression,
\begin{equation}
\ln
\biggl( \frac{T_{c0}}{T_c} \biggl)=\frac{1}{4}\sum_{\alpha=1,2}
\sum_{\sigma=\pm 1}
\sum_{n=0}^{\infty}\frac{v_F^2(q-q_{\alpha\sigma})^2/(4\pi T_c)^2}
{(n+\frac{1}{2})\big[(n+\frac{1}{2})^2+v_F^2(q-q_{\alpha\sigma})^2/(4\pi
T_c)^2\big]},
\end{equation}
where $q_{\alpha\sigma}$ is given by 
Eq. (3). 
Eq.(14) may be rewritten using a so-called $\psi$-function [26],
\begin{equation}
\ln
\biggl( \frac{T_{c0}}{T_c} \biggl)=
\frac{1}{4} 
\sum_{\alpha=1,2}
\sum_{\sigma=\pm 1}
 \biggl( \frac{1}{2} 
 \psi \biggl[  \frac{1}{2} + i \frac{v_F (q-q_{\alpha\sigma})}{4 \pi T_c}
\biggl]
+  \frac{1}{2} \psi \biggl[  \frac{1}{2} - i \frac{v_F (q-q_{\alpha\sigma})}{4 \pi T_c}
\biggl]
- \psi \biggl[ \frac{1}{2} \biggl] 
\biggl) \ .
\end{equation}
[Note that Eqs.(14),(15) are the main analytical results of the Letter.
They connect a transition temperature of the CDW phase, $T_c$, in
the presence of a magnetic field, $H\neq0$, with a transition temperature, 
$T_{c0}$, corresponding to $H=0$ and
$\Delta p=0$.
As it directly seen from Eqs.(14),(15), there exist a competition 
between four nesting vectors, $Q_{\alpha \sigma} = 2 p_F + 
q_{\alpha \sigma}$, from Eq.(3) 
(see Fig.1)]. 

In Fig.3, we present the results of our numerical solutions of 
Eq.(14), which demonstrate a stabilization of the Peierls phase 
with $ Q = 2p_F$ at high enough magnetic fields,
$29 \ T < H < 49 \ T$.
At very high magnetic fields, $H > 49 \ T$, and low magnetic fields,
$H < 29 \ T$, a unique SWS phase is shown to be a ground state 
(see Fig.2).
Note that, in the vicinity of the metal-CDW phases transition line,
the SWS phase is characterized by the following order parameter,
\begin{equation}
\Delta_{SWS}(x) = \cos (qx) \cos(2 p_F x) \ ,
\end{equation}
which corresponds to mixing of two order parameters (7) with
$+q$ and $-q$, where 
$q \neq 0$ [27].
It is important that the SWS phase is characterized by periodically
arranged soliton and anti-soliton walls, where the distance between
them is $x_H =\pi / q$ [21].
In our case, $x_H$ demonstrate a non-trivial dependence on a 
magnetic field.
As it seen from Fig.3, the calculated phase diagram is in good 
qualitative and quantitative agreements with the measured ones 
[13-15].
[Note that for the numerical calculations we have used 
$T_{c} (H=0) = 7 \ K$ [13] and $\Delta p v_F = 60 \ K$.
The latter parameter is in a qualitative agreement with Refs.[17-19]
(see discussions in Ref. [29])].

To summarize, our theory suggests an explanation of the existence 
of the high resistance high magnetic SDW phase in (Per)$_2$Pt(mnt)$_2$ 
conductor [13-15] in terms of the SINe 
effects.
It also predicts the existence of phase transitions between the 
high resistance Peierls phase with large activation gap, $\Delta_p$, 
and a unique SWS phase with two equal magnetic field dependent 
energy gaps, 
$\Delta_{SWS}$.
The SWS phase is also characterized by an activation behavior
of a resistivity with the activation gap being
$\Delta_{SWS} \ll \Delta_p$
(see Fig.2).
It is important that these results are in agreement with the
existing measurements of the activation gaps 
in (Per)$_2$Pt(mnt)$_2$ [13].
In our notations, the measured gaps are: $\Delta_p = 40 \ K \gg
 \Delta_{SWS} \simeq 6-15 \ K$, which are in accordance with the 
 theory.
 We suggest more detailed measurements of the activation
 behavior of resistivity in the vicinities of the Peierls-SWS phase
 transitions to establish the existence of the 
 SWS phase.
 We also think that ac infrared measurements may be useful to 
 detect the existence of two gaps in an electron spectrum of the SWS 
 phase.

We stress that a restoration of the high resistance phase
in a sister compound (Per)$_2$Au(mnt)$_2$ occurs only 
in a magnetic field, perpendicular to the conducting chains, 
where the FIDC effects are
expected [2,4].
On the other hand, its transition temperature is too high to be
explained only by the FIDC effects, which are expected in
the CDW phases only at very low
temperatures [4].
Therefore, for explanation of the phase diagram in
 (Per)$_2$Au(mnt)$_2$ a combination of the FIDC
 and SINe effects has to be considered 
 [28,30].
 In particular, at very high magnetic fields, we expect the
 appearance of the SWS phase in 
 this compound [28].

In conclusion, we point out that a possibility of the SWS phases
to exist in quasi-low-dimensional conductors was earlier discussed
in Refs. [24,31] for very different physical
situations.
To the best of our knowledge, the SWS phase has never been 
experimentally observed yet and, as we hope, it is discovered in 
Q1D (Per)$_2$Pt(mnt)$_2$
conductor.
In particular, the SWS phase does not exist in other CDW systems -
$\alpha$-(ET)$_2$M(SCN)$_4$ materials [5,6] - where corrugations of
the Q1D FS are large and an activation  behavior of resistivity 
is not observed.

One of us (A.G.L.) is thankful to N.N. Bagmet, J.S Brooks, and N. Harrison
for very useful discussions.

$^*$Also Landau Institute for Theoretical Physics,
2 Kosygina Street, Moscow, Russia.

\pagebreak

\begin{figure}[h]
\includegraphics[width=6.5in,clip]{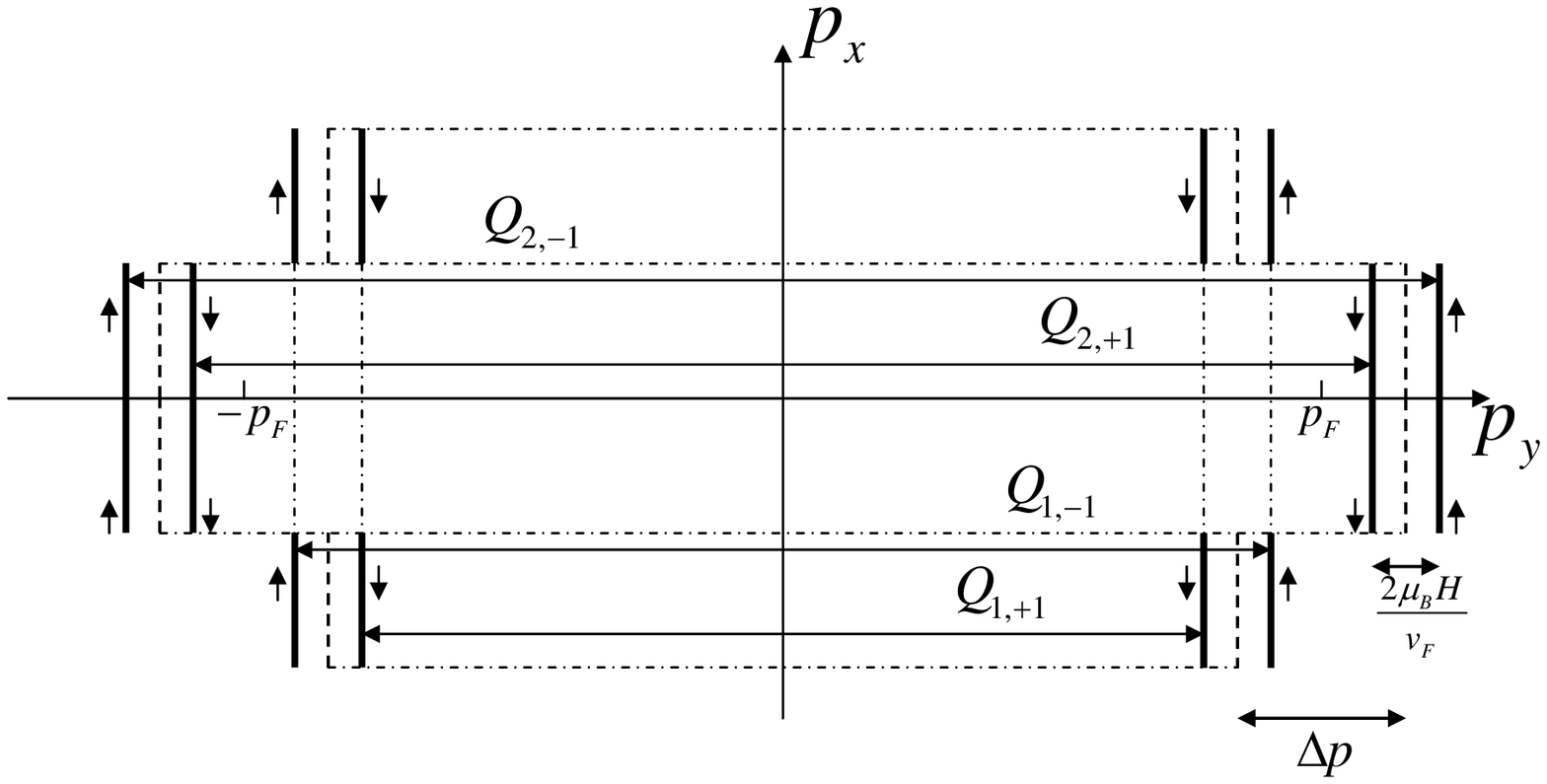}
\caption{
Electron spectrum of a two chains Q1D conductor [16] in a magnetic
field is split into eight sheets of the Fermi surface [see Eq.(2)].
Therefore, there exist a competition between the CDW phases,
characterized by four different nesting vectors, $Q_{1,+1}$,
$Q_{1,-1}$, $Q_{2,+1}$, and $Q_{2,-1}$  
[see Eqs.(3),(6) and 
the text].
At magnetic field $ H^*= \Delta p v_F / 2 \mu_B$, two nesting
vectors coincide, $Q_{1,-1} = Q_{2,+1} = 2 p_F$, which results in 
a restoration of the Peierls CDW phase with $Q = 2p_F$ at high 
magnetic fields [see Eqs.(5),(14) 
and Figs.2,3].
 }
\label{fig1}
\end{figure}

\begin{figure}[h]
\includegraphics[width=5in,clip]{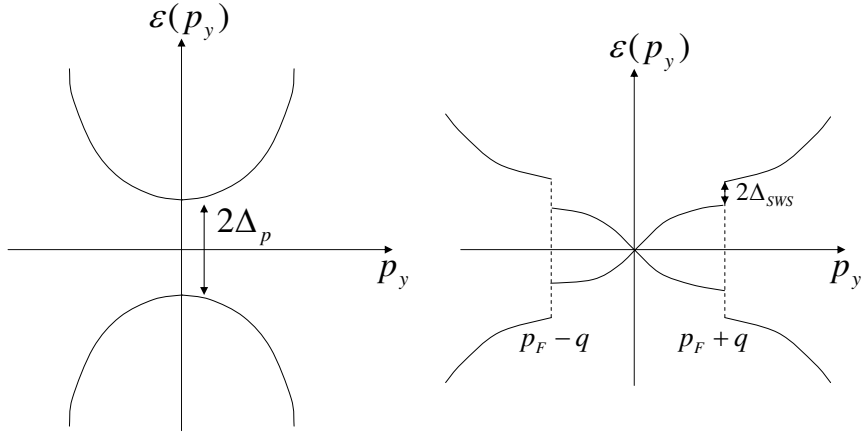}
\caption{
Electron spectrum of the SWS phase with two energy gaps 
(right) is qualitatively different from that in the Peierls phase 
with one energy gap (left) [20] [see Eqs.(5),(16)].
 }
\label{fig1}
\end{figure}

\begin{figure}[h]
\includegraphics[width=5in,clip]{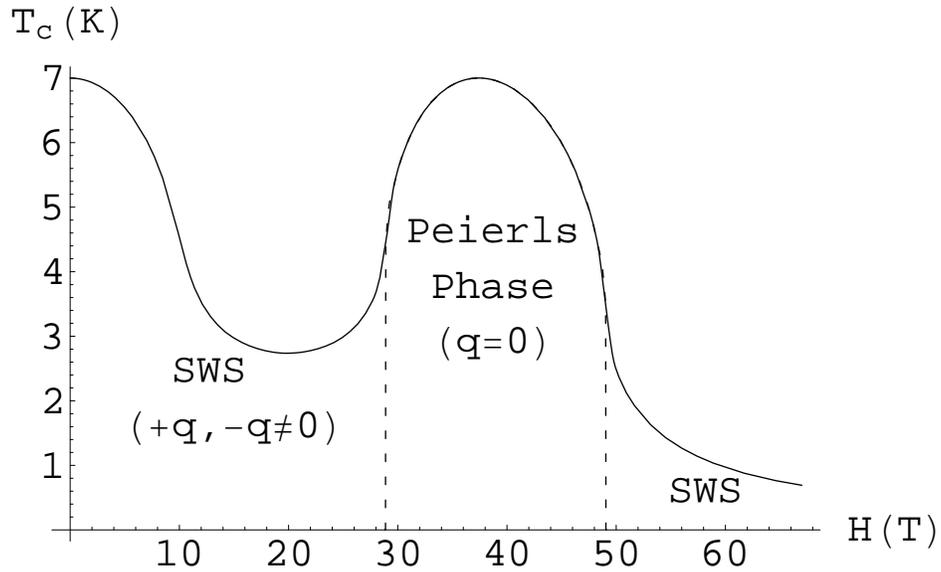}
\caption{
Solid line: phase transition line between the metallic and 
CDW Peierls and SWS phases is calculated by means of Eq.(14).
Dotted lines: phase transitions between the
Peierls and SWS phases [see Eq.(16)].
In our case, $H^* \simeq 40 \ T$ [see Eq.(4)].
 }
\label{fig1}
\end{figure}

\pagebreak


\begin{references}
\bibitem{Lebed-1}
In fact, there exist eight almost plane sheets of the FS
in (Per)$_2$M(mnt)$_2$ conductors, but first four
of them are almost identical to the other four 
(see Ref.[17]) and, thus, we do not distinguish between
them.
\bibitem{Note-1}
We stress that our model is equivalent to that in Ref.[21]
in two limiting cases: at zero magnetic field, $ \mu_B H
/ \Delta p v_F \rightarrow 0$, and at high magnetic fields, 
$ \mu_B H / \Delta p v_F \rightarrow \infty$.
Therefore, at high enough and low enough magnetic fields,
the SWS phase is characterized by two energy gaps at any
temperatures in accordance with Ref.[21] 
(see Fig.2).
As shown in the Letter, the SWS phase is characterized by two energy 
gaps in the vicinity of the metal-SWS phase transition line for the
values of the parameters, $T_c (H=0) = 7 \ K$ and $\Delta p v_F=
60 \ K$. 
Note that, for higher value of the parameter $\Delta p v_F$,
the SWS phase may be characterized by four gaps in its
electron spectrum.
\bibitem{Lebed-5} 
To demonstrate that the periodic SWS phase is a ground
state, we prove that:
(1) two degenerate solutions with $Q_{1,2} = 2p_F \pm q \ (q \neq 0)$
correspond to a maximum of the transition temperature in Eq.(14),
(2) the fourth order terms in the Landau theory of the second 
order phase transitions, in our case, result in mixing of the
above mentioned two order parameters.
As a result, the order parameter (16) is a ground state in the vicinity 
of the metallic-CDW phases transition line, which is known to be a 
corresponding limiting case for the periodic 
SWS [21,24].  
A detailed consideration of the fourth order terms in the Landau
theory will be published elsewhere [28].
\bibitem{Lebed-7} 
From band structure calculations [17], it follows that 
$\Delta p v_F \simeq 35-70 \ K$, an analysis of the quantum interference 
oscillations in (Per)$_2$Au(mnt)$_2$ under pressure gives 
$\Delta p v_F \simeq 70 \ K$, whereas the Landau levels quantization 
analysis  in (Per)$_2$Au(mnt)$_2$ 
gives 
$\Delta p v_F \simeq 50 \ K$.
\bibitem{Lebed-8} 
According to our point of view, a sister compound 
(Per)$_2$Au(mnt)$_2$ is characterized by a high value of 
warping of its Q1D electron spectrum. 
This makes the orbital FIDC effects [2,4] to be as important as the SINe 
ones.
Therefore, a combination of the FIDC and SINe effects restores the 
Peierls CDW phase only for favorite directions of a magnetic field.
\end{references}
\end{document}